\documentclass{ws-procs975x65}

\begin{document}

\title{QUANTUM MECHANICS AND FIELD THEORY WITH MOMENTUM DEFINED  ON AN ANTI-DE-SITTER SPACE}
\author{MYRON BANDER}
\address{Department of Physics and Astronomy\\University of California\\Irvine, CA 92717, USA\\
E-Mail:  mbander@uci.edu}
\begin{abstract}
Relativistic dynamics with energy and momentum resricted to an anti-de-Sitter space is presented, specifically in the introduction of coordiate operators conjugate to such momenta. Definition of functions of these operators, their differentiation and integration, all necessary for the development of dynamics is presented.  The resulting algebra differs from the standard Heisenberg one, notably in that the space-time coordinates do not commute among each other. The resulting time variable is discrete and the limit to continuous time presents difficulties. A parallel approach, in which an overlap function, between position and momentum states, is obtained from solutions of wave equations on this curved space are also investigated. This approach, likewise, has problems in the that high energy behavior of these overlap functions precludes a space-time definition of action functionals. 
\end{abstract}
\bodymatter
\section{Introduction}\label{intro}
Dynamics on space-time manifolds more general than flat Minkowski space leads to restrictions on the corresponding momentum space. For example, placing  coordinate space on a periodic lattice forces momenta to  a hyper-torus $S_1\times \cdots\times S_1$. In general such a construction breaks Lorentz symmetry. In this work we will pursue an opposite approach. We consider energy and momenta to be defined on a space whose isometries include the Lorentz group and in turn investigate the properties of the corresponding position operators.  Specifically, we consider energy and momenta defined on an anti-de-Sitter (AdS) space \cite{bander1}. The full isometry group is $O(2,3)$ which, manifestly contains the $O(1,3)$ Lorentz group.  The group $O(2,3)$ takes the place of the Poincar\'{e} group, the isometry group of Minkowski space.  We loose translation invariance in return for invariance under four additional boost-like transformations. 

The problem now becomes one of identifying the corresponding coordinates. Two approaches are pursued, both based on analogies of relations between time-space and energy-momentum in Minkowski space; both approaches have problems preventing further development.  In the first approach we note that in flat space the coordinates $(t,{\vec x})  $ are operators that translate momenta; as in going to AdS space we lost translation invariance we relate the position operators to the four, aforementioned boosts. The resulting commutation relations among the eight position and momentum operators differ from the Heisenberg algebra, especially in that the position operators do not commute among themselves. Functions of such operators can be introduced as an we can define differentiation of these.  Although it is not obvious how to introduce integration over functions of noncommuting operators, it can be done and the resulting integrals have desired properties. With differentiation and integration procedures in place we are able to define an action integral for a dynamical system. Some of the consequences of this formulation are:
\begin{itemlist}
\item A lower limit on the localizabilty of wave packets,
\item An upper limit on possible masses of particles.
\item Time, instead of being continuous, is discrete. 
\end{itemlist}
In the limit where the curvature of the AdS space goes to zero we recover Minkowski dynamics, with one exception. With $n$ being the discrete time, there are states whose time evolution approaches $(-1)^n\exp(-iEt)$ and thus do not have a reasonable continuum limit. {\em A resolution of this problem is lacking\/}.

In the second approach states with definite time and position are defined via the overlap function $\langle t,{\vec x}|p_0,{\vec p}\rangle$. In flat space this function, $\exp(-ip_\mu x^\mu$ is a solution of the wave equation on momentum space with the coordinates labeling these solutions. We try the same approach of for momenta in AdS space. The problem that arises in this approach is that high energy and momentum behavior of these overlap functions precludes the definition of a position space action.  We cannot even obtain a position space wave equation  corresponding to $p_\mu p^\mu-m^2=0$. Again this problem is unresolved.

Restricting discussion to nonrelativistic dynamics, in which the energy (and time) are treated separatley results in only the spacial momenta being treated as operators. iscussion of this and a proof of the limit of the size of wave packets is presented in the Appendix.
\section{Geometry of anti-de-Sitter Space}\label{geometry}
We consider four dimensional energy-momentum ($p_0,{\vec p}$) on  a an anti-d-Sitter (AdS) hyper-surface embedded in a flat five dimensional Minkowski  space ($p_\tau,p_0,{\vec p}$) subject to the constraint
\begin{equation}\label{ads_condition}
p_\tau^2+p_0^2-{\vec p}\cdot{\vec p}=M^2\, .
\end{equation}
\begin{figure}
\begin{center}
\psfig{file=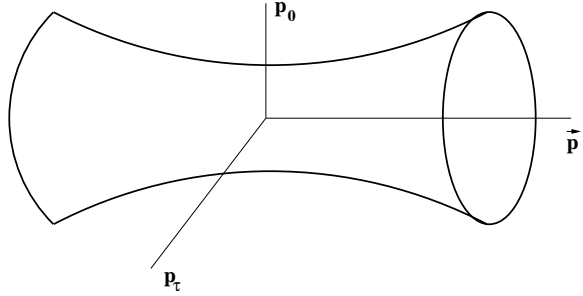, width=3in}
\caption{AdS space embedded in 5-d Minkowski space. he line marked ${\vec p}$ represents three dimensional momenta.}
\label{AdS-space}
\end{center}
\end{figure}
It is convenient to describe this surface using coordinates  ${\vec p}$ and $\omega$ related to $p_\tau$ and to $p_0$ by
\begin{eqnarray}\label{invcoord}
p_\tau&=&\sqrt{{\vec p}\cdot{\vec p}+M^2}\cos\omega\, ,\nonumber\\
p_0&=&\sqrt{{\vec p}\cdot{\vec p}+M^2}\sin\omega\, .
\end{eqnarray}
With these coordinates the invariant energy-momentum volume on this AdS space takes a simple form
\begin{equation}\label{volelem}
d^5p\delta(p^2-M^2)=d{\vec p}d\omega\, ;
\end{equation}
This parametrization makes it is easy to note  that energy-momentum restricted to this curved space places a limit on the mass of any state
\begin{equation}									
m^2=p_0^2-{\vec p}\cdot{\vec p}=M^2\sin^2\omega- ({\vec p}\cdot{\vec p})\cos^2\omega\, ,
\end{equation}
and thus is always less than $M^2$. There is no bound on the energy or momenta, only on the mass of any state. 

\section{Space coordinates}\label{spacecoord}
Placing energy and mo meta on an AdS manifold raises the question of  how to introduce space and time operators. We do this in analogy with a procedure valid for the ordinary situation of energy-momenta in a flat Minkowski space. In that case the  full isometry group of the energy-momentum manifold (not necessarily of any dynamical equations) is the Poincar\'{e} group consisting of the Lorentz transformations and translations generated by 
\begin{equation}\label{minkow-trans}
x^\mu=-i\frac{\partial}{\partial p_\mu}\, ;
\end{equation}
in the present case the full isometry group is the five dimensional anti-de-Sitter group consisting, in addition to Lorentz transformations of the $(p_0\, ,{\vec p})$ subspace, the four Lorentz transformations connecting $p_\tau$ and $\vec p$
\begin{equation}\label{ads-trans}
K^\mu=\sqrt{p_\tau}[-i\frac{\partial}{\partial p_\mu}]\sqrt{p_\tau}\, .
\end{equation}
Using Eq.(\ref{ads_condition}) to write $p_\tau$ in terms of $\vec p$ and  $p_0$, it is straightforward to check that the $SO(3,1)$ Lorentz operators $M^{\mu\nu}=i(p^\mu\frac{\partial}{\partial p\nu}-p^\nu\frac{\partial}{\partial p_\mu})$ and the $K^\mu$'s generate the desired $O(2,3)$ AdS group with the commutation relations
\begin{equation}\label{adscomm}
[K^\mu, K^\nu]=-iM^{\mu\nu}\, .
\end{equation}
The minus sign in the above is crucial as it distinguishes this set of operators as forming the algebra of the aforementioned $O(2,3)$ group, rather than the $O(1,4)$ group that a plus sign would have yielded.

In analogy with Eq.(\ref{minkow-trans}) we postulate the following space-time operators, 
\begin{equation}\label{adsposition}
X^\mu=\frac{K^\mu}{M}\, .
\end{equation}
Again replacing  $p_\tau$ by $(M^2+{\vec p}\cdot{\vec p}-p_0^2)^{\frac{1}{2}}$, we obtain the following coordinate operators:
\begin{equation}\label{coord-oper}
X^\mu=\frac{1}{M}(M^2+{\vec p}\cdot{\vec p}-p_0^2)^{\frac{1}{4}}\left(-i\frac{\partial}{\partial p_\mu}\right)(M^2+{\vec p}\cdot{\vec p}-p_0^2))^{\frac{1}{4}}\, .
\end{equation}
In the limit $M\rightarrow\infty$ these position operators go over to the usual ones in Eq.(\ref{minkow-trans}).  It is amusing to note that the identification in Eq.(\ref{adsposition}) together with the  commutation relation in Eq.(\ref{adscomm}) reproduces the spacial noncommuting quantum mechanics originally introduced by Snyder\cite{Snyder:1946qz} in 1946. 
\section{Hilbert space and Modified Heisenberg Algebra}
We shall be working primarily in the  Hilbert space consisting of eigenststes of the operators $p_0$ and $\vec p$, labeled as$|p_0,{\vec p}\rangle$. Using the parametrization of  Eq.(\ref{invcoord}) the inner product of these states is 
\begin{equation}\label{innprod}
\langle p_0',{\vec p}'|p_0,{\vec p}\rangle= \delta(\omega'-\omega)\delta({\vec p}-{\vec p}'), .
\end{equation}

The Heisenberg algebra of momenta and the coordinate operators defined in Eq.(\ref{coord-oper}) is modified from the usual one to
\begin{eqnarray}
[p_\mu, X_\nu]&=&ig_{\mu\nu}\frac{p_\tau}{M}\, , \nonumber \\
{[X_\mu, X_\nu]}&=& -i\frac{M_{\mu\nu}}{M^2}\, . 
\label{H-alg}
\end{eqnarray}
Again, in the limit $M\rightarrow\infty$ we recover the usual commutation relations.  As mentioned earlier, the space-space commutator is the one discussed in Ref. [\refcite{Snyder:1946qz}].

\section{Functions of position operators and differentiation of these}
A function $f(X_\mu)$ of the operators introduced in Eq.(\ref{adsposition}) corresponding to one of the ordinary position operators, $f(x_\mu)$ can be obtained by assuming they have the same Fourier transforms. Namely
\[
f(x_\mu)=\int d^4q {\tilde f(q)}e^{iq\cdot x}
\] 
leads to the suggestion that we define the corresponding $f(X)$ as
\begin{equation}\label{funcX1}
f(X_\mu)=\int d^4q {\tilde f(q)}e^{iq\cdot X}\, .
\end{equation}
However, for technical reasons to which we shall soon return (see discussion towards the end of Sect.(\ref{intfX})), we will modify Eq.(\ref{funcX1}); we first introduce a vector $Q_\mu$ related to $q_\mu$,
\begin{equation}\label{qtoQ}
Q_\mu=\frac{q_\mu}{M}\arcsin\left(\frac{q}{M}\right)\, ,
\end{equation}
The above definition is valid for $q$ timelike; with the arcsin going over to an arcsinh when $q$ is spacelike. We note that for small $q/M$
$Q_\mu\rightarrow q_\mu$. With this definition of $Q_\mu$  change Eq.(\ref{funcX1}) to
\begin{equation}\label{funcX2}
f(X_\mu)=\int d^4q {\tilde f(q)}e^{iQ\cdot X}\, .
\end{equation}

The derivative of $F(X)$  is, as expected, defined as as
\begin{equation}\label{dfX}
\frac{\partial f(X)}{\partial X_\mu}=-i[p_\mu, f(X)]
\end{equation}

\section{Integration of functions of the operators X}\label{intfX}
For many purposes, both in quantum mechanics and in field theory we need to define an ``integral" over the operator $X_\mu$. 
Primarily, we want to be able to define an action whose variation will yield appropriate equations of motion.  With this in mind we will 
abstract from the definition of ordinary integration the steps needed to carry over this procedure to functions, as defined previously, of 
the noncommuting specie-time coordinates, $X_\mu$. 
For ordinary functions we can use the Fourier transforms, ${\tilde f}_i(q_\mu)$ of $f_i(x_\mu)$ to obtain the integral of
\begin{equation}\label{intdef1}
\int d^4x f_1(x)\cdots f_N(x)=(2\pi)^4\int d^4q_1\cdots d^4q_N {\tilde f}_1(q_1)\cdots{\tilde f}_N(q_N\delta^4(q_1+\cdots q_n)\, .
\end{equation}
At this point we are left with the problem of finding the analog of the $\delta$ function in the above valid for our coordinates. 
Noting that the position operator acts as a translation operator on momentum states, $\exp(iq\cdot x)|p\rangle=|p+q\rangle$ allows us to represent the delta function in Eq. (\ref{intdef1}) as
\begin{equation}
\delta^4(q_1+\cdots +q_N)=\langle p_0, {\vec p}|e^{i(q_1+\cdots +q_N)\cdot x}|p_0,{\vec p}\rangle\, ,
\end{equation}
where $|p_0,{\vec p}\rangle$ is any state. Carrying this over to the representation of  functions of $X_\mu$ as given in Eq.(\ref{funcX2}) yields
\begin{eqnarray}
``\int  d^4X"f_1(X)\cdots f_N(X)=(2\pi)^4&\int& d^4q_1\cdots d^4q_N {\tilde f}_1(q_1)\cdots{\tilde f}_N(q_N)\times\nonumber\\
&{}&\langle p_0,{\vec p}|
  e^{iQ_1\cdot X}\cdots e^{iQ_N\cdot X}|p_0, {\vec p}\rangle\, .
\end{eqnarray}
As mentioned, the $(p_0, {\vec p})$ can refer to any state; for most calculations it is convenient to take the above matrix elements in the state $|0,{\vec 0}\rangle$.  the AdS symmetry insures that this definition is independent of the choice of the state $(p_0, {\vec p})$.

Some properties of this integration prescription are:
\begin{enumerate}
\item[(i)]With derivatives defined by Eq.(\ref{dfX}) we find  
\begin{equation}
``\int d^4X"\partial_\mu f(X_\mu)=0\, .\nonumber
\end{equation}
\item[(ii)] \begin{equation}
``\int d^4X" e^{iq^1\cdot X}e^{-iq^2\cdot X}=\delta(q_\mu^1-q_\mu^2)\, .\nonumber
\end{equation}
\end{enumerate}
Had we used $q_\mu$ instead of $Q_\mu$, Eq.(\ref{qtoQ}), in the definition of $f(X)$, Eq.~(\ref{funcX2}), the right hand side of item(ii), above, would have been multiplied by $q/[M\sin(q/M))$.

\section{Translation Invariance, or lack thereof}\label{transinv}
The modified Heisenberg algebra, Eq.~(\ref{H-alg}) precludes having a unitary operator  shifting the position operator $X$.
Using the momentum operator produces
\begin{equation}\label{Ptrans}
e^{ip\cdot a}X_\mu e^{-ip\cdot a}= X_\mu+a_\mu\times\frac{p_\tau}{M}\, .
\end{equation}
This should come as no surprise as the isometry group of our space is the de Sitter group consisting of Lorentz transformations and the four boosts $K_\mu$, Eq.~(\ref{ads-trans}, involving the $\tau$ direction, and not the Poincar\'{e} group consisting of Lorentz transformations and translations. 

The fact that the different components of the position operators do not commute puts a limit on localizing wave packets. It is straightforward to show, see Appendix A,  that the expectation value of $X_1^2+X_2^2+X_3^2$ in any packet must exceed $1/M^2$.

Integrals of products of more than two fields require the evaluation of matrix elements of the form $\langle 0,{\vec 0}|exp(iQ_1\cdot X)\cdots |exp(iQ_N\cdot X)| 0,{\vec 0}\rangle$; the results are complicated and no closed expression is available.  The order of the exponentials cannot, in general, be reversed; this is another indication of the noncommutativity of the operators $X_\mu$. 

\section{Discreteness of Time}
We may diagonalize one of the space-time coordinates and we choose it to be $X_0$. In the $\omega,{\vec p}$ parametrization,Eq.~(\ref{invcoord}), it takes a simple form
\begin{equation}\label{timecoord}
X_0=\frac{-i}{M}\frac{\partial}{\partial\omega}\, ;
\end{equation}
the eigenvalues of this time variable are discrete, $t=n/M$, with $n$ integer. In the $M\rightarrow\infty$ limit time goes over to a continuum limit.  With time discrete we expect the energy interval to be finite for a fixed $\vec p$ and indeed Eq.~(\ref{invcoord}) shows that $|E|\le\sqrt{{\vec p}\cdot{\vec p}+M^2}$. Parametrizing $\omega$ as $\omega=E/M$ leads to, in the large $M$ limit, the identification $p_0=E$. Subsequently we will encounter problems with this interpretation.

\section{Field Theory}{\label{field_theory}
We shall try a naive procedure to set up a field theory where the fields $\phi(X)$ are functions of the operators $X_\mu$ by postulating action functionals for these fields. For a free field with mass $\mu$ action is taken to be
\begin{eqnarray}\label{freefield}
S_F[\phi(X)]&=&``\int d^4X" \left\{-[p_\nu,\phi^\dag(X)][p^\nu,\phi(X)]-\mu^2\phi^\dag(X) \phi(X)\right\}\nonumber\\
  &=&\langle 0{\vec 0}| -[p_\nu,\phi^\dag(X)][p^\nu,\phi(X)]-\mu^2\phi^\dag(X) \phi(X)|0,{\vec 0}\rangle\, .
\end{eqnarray}
For $\phi(X)$ of the form $\phi(X)=\int d^3p{\tilde \phi}(q)\exp[iQ\cdot X]$ we readily obtain the mass shell condition $q^2-\mu^2=0$ ($q$ and $Q$ are related by Eq.~(\ref{qtoQ})) Parametrizing $q_\mu$ as in Eq(\ref{invcoord}), namely,  $q_0=\sqrt{{\vec q}\cdot{\vec q}+M^2}\sin\omega$, this mass condition translates to
\begin{equation}
\sin\omega=\pm\frac{\sqrt{{\vec p}\cdot{\vec p}+\mu^2}}{X}
\end{equation}
For $M$ large we obtain four solutions
\begin{eqnarray}\label{omegasolts}
\omega&=&\pm\sqrt{{\vec p}\cdot{\vec p}+\mu^2}/M\, ,\nonumber\\
&{}&\\
\omega&=&\pm [\pi -\sqrt{{\vec p}\cdot{\vec p}+\mu^2}/M]\, .\nonumber
\end{eqnarray}

As the time evolution of the states is 
\begin{equation}\label{time_evol}
|S;t=n\rangle=|S;t=0\rangle e^{i\omega n}\, ,
\end{equation}
we have to  interpret these four solutions.
The $\pm$ indicates the usual positive and negative frequency solutions; the first solution goes over to the usual time dependence $\exp(-iEt)$ while the second one has a discrete time propagation of the form $(-1)^n\exp(-iEt)$ which has no smooth continuum limit. {\it An interpretation of this behavior is lacking at present and this is one of the problems we mentioned in the Introduction.\/}

\section{Space-time Coordinates}\label{spacetime}
In order to resolve the problem brought up at the end of the last section and for possible computational simplifications we shall try a different method of introducing coordinates appropriate to momenta on an AdS space. Again we shall use analogies with such procedures in flat Minkowski space as a guide. This time, however, rather then studying space-time operators we will look for states that correspond, in the $M\rightarrow\infty$ limit to the usual ones, $|t, {\vec x}\rangle$. As the $X_\mu$'s do not commute, we cannot look for simultaneous eigenstates of these. For flat energy-momentum we can relate momentum and position eigenstates by the overlap
\begin{equation}\label{flatoverlap}
\langle x_0,{\vec x}|p_0,{\vec p}\rangle=\frac{1}{(2\pi)^2}e^{i{\vec p}\cdot{\vec x}}\, .
\end{equation}
The states $|t,{\vec x}\rangle$ are eigenstates of the commuting operators $x_\mu$. As mentioned earlier, in the present situation with the operators $X_\mu$  not commuting we cannot define the analogous state $|X_0, {\vec X}\rangle$. The question we will address in this section is whether we can still obtain a version of the right hand side of Eq.~(\ref{flatoverlap}) and thus define a state $|X_0, {\vec X}\rangle$ as $\int d^4p |p_0,{\vec p}\rangle\langle p_o,{\vec p}||X_0, {\vec X}\rangle$.  

We note that Eq.~({\ref{flatoverlap}) or its spherical coordinate version,
\begin{equation}\label{flatoverlap2}
\langle x_0; x,\theta,\phi|p_0;p,l,m\rangle =e^{-ip_0x_0}\frac{1}{\sqrt{2\pi}}j_l(pr)Y_{l,m}(\theta,\phi)\, ,
\end{equation} 
are eigenfunctions of the wave equation on momentum space with $(x_0,x,l,m)$ or $(x_0,x,\theta, \phi)$  labeling the solutions,
\begin{equation}\label{flatwave}
[\partial_{p_0}\partial_{p_0}-\partial_{p_i}\partial_{p_i}]e^{-ip_0x_0}j_l(px)Y_{lm}(\theta,\phi)=(-x_0^2+x^2)e^{-ip_0x_0}j_l(px)Y_{lm}(\theta,\phi)\, .
\end{equation}

We are thus lead to look for eigenfunctions of 
\begin{equation}\label{dSwave}
g^{\mu\nu}\frac{\partial}{\partial p_\mu}\frac{\partial}{\partial p_\nu}=K_\mu K_\mu-M_{\mu\nu}M^{\mu\nu}\, .
\end{equation}
$g^{\mu\nu}$ is the AdS metric, $ds^2=({\vec p}\cdot{\vec p}+M^2)d\omega^2-{d\vec p}\cdot{d\vec p}\, ;$ The right hand side in Eq.~(\ref{dSwave}) is the Casimir operator for the anti-de-Sitter group.

Explicitly this wave equation ,written using the energy-momentum coordinates ${\vec q}={\vec p}/M$ and$\omega$,   is
\begin{eqnarray}\label{dSwave2}
-g^{\mu\nu}\frac{\partial}{\partial p_\mu}\frac{\partial}{\partial p_\nu}  &=&\frac{\partial}{\partial q}(1+q^2)\frac{\partial}{\partial q}+(\frac{2}{q}+4q)\frac{\partial}{\partial q}-
\frac{{\vec L}^2}{q^2}-\frac{1}{1+q^2}\frac{\partial^2}{\partial\omega^2}\nonumber\\
&=&M^2x^2\, .
\end{eqnarray}
The solutions to the above are 
\begin{equation}\label{wavefunc}
Z_{\lambda,l,m:n}(q, {\hat q}; \omega)=B^{l,n}_{-\frac{1}{2}+i\lambda}(iq)Y_{l,m}(\hat q)e^{-in\omega}\, ,
\end{equation}
where the functions $B^{l,n}_{-\frac{1}{2}+i\lambda}(iq)$ are related to the Gegenbauer polynomials\cite{gradshteyn} and the parameter $\lambda=M^2 x\cdot x$ can be real, implying space-like $x$ or equal to $iN$, with $N\le (n-1)$ integer for time-like $x$.  

Summarizing these results we shall try to  obtain a local field theory for states $|x_\mu\rangle$ related to the momentum ones by 
\begin{equation}\label{dSoverlap}
\langle n; x,l,m|\omega, {\vec p}\rangle=Z_{\lambda,l,m:n}(Mq, {\hat q}; \omega)\, .
\end{equation}
In the limit $M\rightarrow\infty$ Eq.~(\ref{dSoverlap}) approaches Eq.~(\ref{flatoverlap}).

\subsection{Problem}\label{problem}
We would like to investigate analog of the Klein-Gordon or  the free field equation in the states defined by Eq.~(\ref{dSoverlap}),
\begin{equation}
\langle n(1q'; x',l',m'|\left[({\vec p}\cdot{\vec p}+M^2)\sin^2\omega-({\vec p}\cdot{\vec p}\right]|x_0; x,l,m\rangle\, .
\end{equation}
{\it The problem is that due to the large q behavior of the functions $B^{l,n}_{-\frac{1}{2}+i\lambda}(iq)$, Eq.~(\ref{wavefunc}), these matrix elements do not converge, not even to Dirac delta functions or to their derivatives\/}. The easiest way to see this problem is to look in detail at the case of $(1+1)$ dimensions where Eq.~(\ref{dSwave2}) takes on a simpler form
\begin{equation}\label{dSwave3}
-g^{\mu\nu}\frac{\partial}{\partial p_\mu}\frac{\partial}{\partial p_\nu}[{\rm (1+1) dim}]=
\frac{\partial}{\partial q}(1+q^2)\frac{\partial}{\partial q}-\frac{1}{1+q^2}\frac{\partial^2}{\partial\omega^2}\, .
\end{equation}
The eigenfunctions of the above are Legendre functions \cite{gradshteyn} of imaginary argument; smooth behavior at $q=0$ restricts them to the form
\begin{equation}\label{2doverlap}
\langle n; x|\omega;Mq\rangle=\left\{\begin{array}{c}
        P_{-\frac{1}{2}+i\lambda}^n(iq)\, ; \mbox{$M^2x^2=\lambda^2-\frac{1}{4}$; x space-like ,}\\\
        Q_{-\frac{1}{2}-N}^n(iq)\, \mbox{$N\le(n-1)\, ; M^2x^2=N^2+\frac{1}{4}$; x timelike.}
        \end{array}\right.
        \end{equation}
        As the large $q$ behavior of $P_{-\frac{1}{2}+i\lambda}^n(iq)$ is $q^{-\frac{1}{2}+i\lambda}$, the matrix elements of $q$, $q^2$, etc. do not converge.
Again, the resolution of this problem is unclear.
\section*{Acknowledgements}
This article is based on a talk at a symposium celebrating the 80th birthday of Professor Murray Gell-Mann held at Nanyang Technical University, Sinapore, 24th – 26th February, 2010.  I wish to thank Professor Harald Fritzsch for organizing this meeting and for inviting me to give this talk.

\appendix{Nonrelativistic (three dimensional) quantum mechanics}
A simpler application of the ideas discussed in this article can be used to study the case where only the three momenta are placed on a de Sitter space, which in this case may be viewed as a surface embedded in a (3+1) dimensional Minkowski space with coordinates ($p_\tau,{\vec p}$) subject to the constraint 
\begin{equation}\label{3-d-constraint}
p_\tau^2-{\vec p}\cdot{\vec p}=M^2\, ;
\end{equation}
The energy coordinate, $p_0$ ranges over the full interval, $-\infty\le p_0 \le\infty$. The operators conjugate to $p_0, {\vec p}$ are $t$ and, using eq.~(\ref{ads-trans})  and (\ref{adsposition}) as a guide,  
\begin{equation} \label{3-d-position}
X_i=\frac{1}{M}({\vec p}\cdot{\vec p}+M^2)^{\frac{1}{4}}\left(i\frac{\partial}{\partial p_i}\right) ({\vec p}\cdot{\vec p}+M^2)^{\frac{1}{4}}\, .
\end{equation}
This time The Heisenberg algebra is modified to
\begin{eqnarray}\label{3-d-Heisalg}
[p_i, X_j]&=&-i\delta_{ij}\frac{p_0}{M}\, .\nonumber\\
&{}&\\
\left[X_i, X_j\right] &=&-i\frac{M_{ij}}{M}\, ,\nonumber
\end{eqnarray}
where $M_{ij}$ is the angular momentum. Again the minus sign in front of the $M_{ij}$ is crucial to distinguish this as the (1,3) Lorentz group rather than the SO(4) rotation group. 

As expected, the noncommutativity of the X's prevents a localization of wave packets. The extent to which a packet may be localized is controlled by the eigenvalues of $X^2$ and a lower bound on such eigenvalues may be obtained by noting that the $SO(1,3)$ Casimir operator ${\cal K}^2 -J^2$ 
equals $\rho^2-j_0^2+1$ \cite{Representation} for representations labeled by $(\rho, j_0)$, with real $\rho\ge 0$, and with all angular momenta in the representation having values greater than $ j_0$. As $X^2=\left({\cal K}^2-J^2+J^2\right)/M^2$, its eigenvalues are $\left[\rho^2+1-j_0^2+j(j+1)\right]/M^2$, with $j\ge j_0$; thus we find that $X^2\ge 1/M^2$ and wave packets cannot be localized to better than $1/M$.

Quantum mechanics for one particle in an external potential $V(X)$ involves the operator eigenvalue equation
\begin{equation}\label{Schr-eq}
E=\frac{1}{m}{\vec p}\cdot{\vec p}+ V(X)\, ,
\end{equation}
while the analogous two body problem requires a bit more care.  Due to noncummutativity of the coordinate components we cannot follow the usual procedure and change coordinates from $X^{(1)}$ and $X^{(2)}$ to relative and center of mass ones; these have to be introduced from the beginning.  With the usual definitions of relative and center of mass coordinates,  $\vec p_{\rm rel}=(m^{(2)}\vec p^{\, (1)}-m^{(1)}\vec p^{\, (2)})/(m^{(1)}+m^{(2)})$, $\vec x_{\rm rel}=\vec x^{\,  (1)}-\vec x^{\, (2)}$ and $\vec p_{\rm cm}=\vec p^{\, (1)}+\vec p^{\, (2)}$, $\vec x_{\rm cm}=(m^{(1)}\vec x^{\, (1)}+m^{(2)}\vec x^{\, (2)})/(m^{(1)}+m^{(2)})$, we define
\begin{eqnarray}\label{cm-rel}
X_i^{\rm rel}&=&\frac{i}{M}\sqrt{\vec p_{\rm rel}\cdot\vec p_{\rm rel}\cdot+M^2}
       \left(\frac{\partial}{\partial p_i^{(1)}}-\frac{\partial}{\partial p_i^{(2)}}\right)
           \sqrt{\vec p_{\rm rel}\cdot\vec p_{\rm rel}\cdot+M^2}\, ,\nonumber\\
&{}&\\
X_i^{\rm cm}&=&\frac{i}{M(m^{(1)}+m^{(2)})}\sqrt{\vec p_{\rm cm}\cdot\vec p_{\rm cm}\cdot+M^2}
       \left(m^{(1)}\frac{\partial}{\partial p_i^{(1)}}+m^{(2)}\frac{\partial}{\partial p_i^{(2)}}\right)
           \sqrt{\vec p_{\rm cm}\cdot\vec p_{\rm cm}\cdot+M^2}\, .\nonumber
\end{eqnarray}
A direct computation shows that these relative and center of mass variables commute and the coordinates within each class obey the commutation relations of eq.~(\ref{3-d-Heisalg}) and have the desired limit for large $M$.  From the start we would formulate a two body problem as
\begin{equation}\label{2body}
H=\frac{p^{\, (1)2}}{2m^{\, (1)}}+\frac{p^{\, (2)2}}{2m^{\, (2)}}+V(X^{\rm rel})\, .
\end{equation}
The use of these relative coordinated may be extended to many body situations.


\begin{thebibliography}{9}
\bibitem{bander1}
Bander1
\bibitem{Snyder:1946qz}
  H.~S.~Snyder,
  Phys.\ Rev.\  {\bf 71},  (1947)  38.
\bibitem{gradshteyn}
Gradshteyn
\bibitem{Representation}
Gelfand
\end{thebibliography}
\end{document}